# Energy-filtered transmission electron microscopy of biological samples on highly transparent carbon nanomembranes

Daniel Rhinow[1*], Matthias Büenfeld[2], Nils-Eike Weber[2], André Beyer[2], Armin Gölzhäuser[2], Werner Kühlbrandt[1], Norbert Hampp[3], Andrey Turchanin[2]

[1] *Max-Planck-Institute of Biophysics, Department of Structural Biology, Max-von-Laue-Str. 3, D-60439 Frankfurt, Germany*

[2] *University of Bielefeld, Department of Physics, D-33615 Bielefeld, Germany*

[3] *University of Marburg. Department of Chemistry, Hans-Meerwein-Straße, D-35032 Marburg, Germany*

\* corresponding author: daniel.rhinow@biophys.mpg.de, Tel.: +49-69-6303-3050

**Abstract**

Ultrathin carbon nanomembranes (CNM) comprising crosslinked biphenyl precursors have been tested as support films for energy-filtered transmission electron microscopy (EFTEM) of biological specimens. Due to their high transparency CNM are ideal substrates for electron energy loss spectroscopy (EELS) and electron spectroscopic imaging (ESI) of stained and unstained biological samples. Virtually background-free elemental maps of tobacco mosaic virus (TMV) and ferritin have been obtained from samples supported by ~ 1 nm thin CNM. Furthermore, we have tested conductive carbon nanomembranes (cCNM) comprising nanocrystalline graphene, obtained by thermal treatment of CNM, as supports for cryoEM of ice-embedded biological samples. We imaged ice-embedded TMV on cCNM and compared the results with images of ice-embedded TMV on conventional carbon film (CC), thus analyzing the gain in contrast for TMV on cCNM in a quantitative manner. In addition we have developed a method for the preparation of vitrified specimens, suspended over the holes of a conventional holey carbon film, while backed by ultrathin cCNM.

# 1. Introduction

Electron cryo-microscopy (cryoEM) of vitrified biological specimens is a powerful method for the analysis of macromolecular structures, ranging from single particles up to tomographic volumes of whole cells [1-4]. Biomolecules are phase objects and image contrast is degraded by radiation damage, inelastic scattering, electrostatic charging, and specimen movement [5]. Among the technical efforts to increase the notoriously low signal-to-noise ratio of EM images of biological macromolecules the development of support films other than the routinely used amorphous carbon has attracted interest of several groups. Various new materials have been tested recently as support films for electron microscopy, among them conductive amorphous TiSi alloys [6] as well as ultrathin carbonaceous substrates like graphene [7], graphene oxide [8,9], and carbon nanomembranes [10-12].

The implementation of energy filters to remove inelastically scattered electrons has considerably improved structural analysis of ice-embedded biological specimens [13,14]. Furthermore, an energy filter converts a conventional electron microscope into a powerful analytical tool for electron energy loss spectroscopy (EELS) and electron spectroscopic imaging (ESI) [15]. Widely used in materials science, there is growing interest in soft matter and biological applications of EELS and ESI [16-20].

In this work we combine the advantages of ~1 nm thin carbon nanomembrane (CNM) support films with the potential of energy-filtered transmission electron microscopy (EFTEM) and demonstrate that CNM are well-suited for background-free mapping of chemical elements within stained and unstained biological specimens. Furthermore, using two different ways of specimen preparation, we demonstrate that conductive carbon nanomembranes (cCNM) are promising support films for cryoEM of ice-embedded specimens.

.

## 2. Materials and methods

*2.1 Fabrication of CNM and cCNM support films*

Ultrathin carbon nanomembranes (CNMs) were fabricated by transferring crosslinked self-assembled monolayers (SAMs) of 1,1′-biphenyl-4-thiol (BPT, Platte Valley Scientific) onto TEM grids. The BPT SAMs were prepared on gold/mica substrates with a thickness of the gold layer of 300 nm (G. Albert PVD, Silz, Germany). The substrates were cleaned in a UV cleaner (UVOH 150 LAB from FHR, Germany) for 5 minutes, rinsed with ethanol and dried in a stream of nitrogen. Afterwards they were immersed in a 1 mM solution of BPT in N,N-dimethylformamide (DMF p.a., Sigma-Aldrich, dried with 0.4 nm molecular sieve) at room temperature (RT) for 3 days. Subsequently the samples were rinsed with DMF, ethanol (p.a., VWR) and finally dried in a stream of nitrogen. This preparation results in densely packed BPT SAMs on Au with a thickness of ~1 nm [21]. The BPT SAMs were crosslinked by electron exposure [22] with a flood-gun (Specs FG20) in high vacuum (<5•$10^{-7}$ mbar). Electron energy of 100 eV and a dose of ~50 mC/cm$^2$ (80 e$^-$/Å$^2$) were applied to achieve complete crosslinking [23] and to transform a SAM into a molecular nanomembrane consisting of disordered aromatic carbon rings. The dose was measured with a Faraday cup in close proximity to the sample.

To fabricate cCNMs the cross-linked BPT SAMs were annealed at high temperatures under UHV conditions. To this end, cross-linked BPT SAMs (nanomembranes) with the underlying gold layer were first cleaved from the mica by immersion in hydrofluoric acid (48%) for 15 min and transferred onto a clean quartz substrate using a spin-coated (500 nm) and baked layer of polymethylmethacrylate (PMMA) for stabilization [11]. The PMMA layer was then dissolved in acetone to yield a clean nanomembrane surface. The samples were annealed at ~1200 K in Mo sample holders with a resistive BN-heater using a heating/cooling rate of ~150 K/h and an annealing time of ~0.5 h. Annealing of the nanomembranes on the Au/quartz substrates results in substantially lower defect density in cCNMs in comparison to the annealing on the original Au/mica substrates [11]. Annealing temperature was controlled with a Ni/Ni-Cr thermocouple and two-color pyrometer (SensorTherm). This procedure transforms insulating CNMs into cCNMs with a sheet resistivity of ~100 kΩ/sq at RT [11]. Accounting for the thickness of cCNM, which was determined to 0.7 nm in former work [11], this corresponds to a bulk resistivity of ~ 7*10exp(-5) Ωm.

Transfer of both CNMs and cCNMs from the gold surfaces onto TEM grids was done by removing a PMMA/nanomembrane/Au stack from the underlying mica or quartz surface as described above, etching gold in an $I_2$/KI-etch bath (15 min) and transferring the nanomembrane/PMMA to TEM grids with the holey carbon or lacey carbon films (Plano). Afterwards the PMMA was dissolved in acetone using a critical point dryer to yield clean nanomembranes. Such a transfer procedure is very effective and results in large-scale and robust nanomembranes with the size of suspended areas more than 200 μm [24, 25].

## 2.2. Conventional TEM of negatively stained TMV on CNM

Tobacco mosaic virus (TMV) was a kind gift of Ruben Diaz-Avalos (New York Structural Biology Centre). TMV was prepared on CNM supports, which had been glow discharged in air for 15 sec, and stained with uranyl acetate. Samples were analyzed at 120 kV accel~erating voltage in a FEI Tecnai Spirit microscope. Images were recorded on a 2k x 2k CCD camera (Gatan) and analyzed with imageJ (Rasband, W.S., ImageJ, U. S. National Institutes of Health, Bethesda, Maryland, USA, http://rsb.info.nih.gov/ij/, 1997-2009). The gain in contrast was quantified by calculating the ratio $I_{background}$-$I_{TMV}$/$I_{background}$ for TMV on CNM as well as on conventional carbon (CC). The average intensities $I_{background}$ and $I_{TMV}$ were measured from boxed areas of 2100 pixels each.

## 2.3. Energy-filtered imaging and ESI of TMV and ferritin on CNM

Horse spleen ferritin was kindly provided by Karen Davies (MPI of Biophysics). TMV and ferritin were prepared on CNM supports, which had been glow discharged in air for 15 sec. TMV was stained with uranyl acetate. Samples were analyzed at liquid nitrogen temperature in a FEI Tecnai Polara microscope equipped with a field emission gun, operating at 300 kV, a post-column energy filter, and a 2k x 2k CCD camera (Gatan). Grids were loaded under liquid nitrogen in cartridges for the multispecimen carrier of the

Polara microscope. Elemental maps were obtained using the three-window method. In the case of negatively stained TMV on CNM energy-selected images were acquired in the vicinity of the uranium O edge (96 eV, slit width 5 eV, 20 s acquisition time) and the carbon K edge (284 eV, slit width 20 eV, 40 s acquisition time). In the case of ferritin on CNM energy-selected images were acquired in the vicinity of the iron M edge (54 eV, slit width 8 eV, 40 s acquisition time). All images were processed by means of Digital Micrograph Software (Gatan).

*2.4. EFTEM of ice-embedded TMV on cCNM*

TMV was a kind gift of Carsten Sachse (EMBL, Heidelberg). 3.5 µl of TMV solution were applied to conductive carbon nanomembranes (cCNM), supported by lacey carbon, which had been glow discharged in air for 60 sec. In one method cCNM were loaded from top, blotted for 2 sec, and subsequently plunged into liquid ethane. In a second approach grids were loaded from the carbon facing side of cCNM. Samples were blotted for 5 s and plunged into liquid ethane. Grids were loaded under liquid nitrogen in cartridges for the multispecimen carrier of the Polara microscope. Zero-loss filtered CCD images of ice-embedded TMV, supported by cCNM only, were obtained at liquid nitrogen temperature in a FEI Tecnai Polara microscope using a 2k x 2k CCD camera (Gatan). Non-filtered images were recorded on a 4k x 4k CCD camera (Gatan). Images were recorded with an underfocus of 3 µm. The gain in contrast was quantified by calculating the ratio $I_{background}$-$I_{TMV}$/$I_{background}$ for ice-embedded TMV on cCNM as well as on CC.

## 3. Results and discussion

*3.1. EFTEM and ESI of biological specimens on CNM*

Ultrathin carbon nanomembranes (CNM) are highly transparent substrates, which considerably improve the contrast of supported biomolecules. Fig. 1a shows an unfiltered image of negatively stained tobacco mosaic virus (TMV) adsorbed to a CNM support.

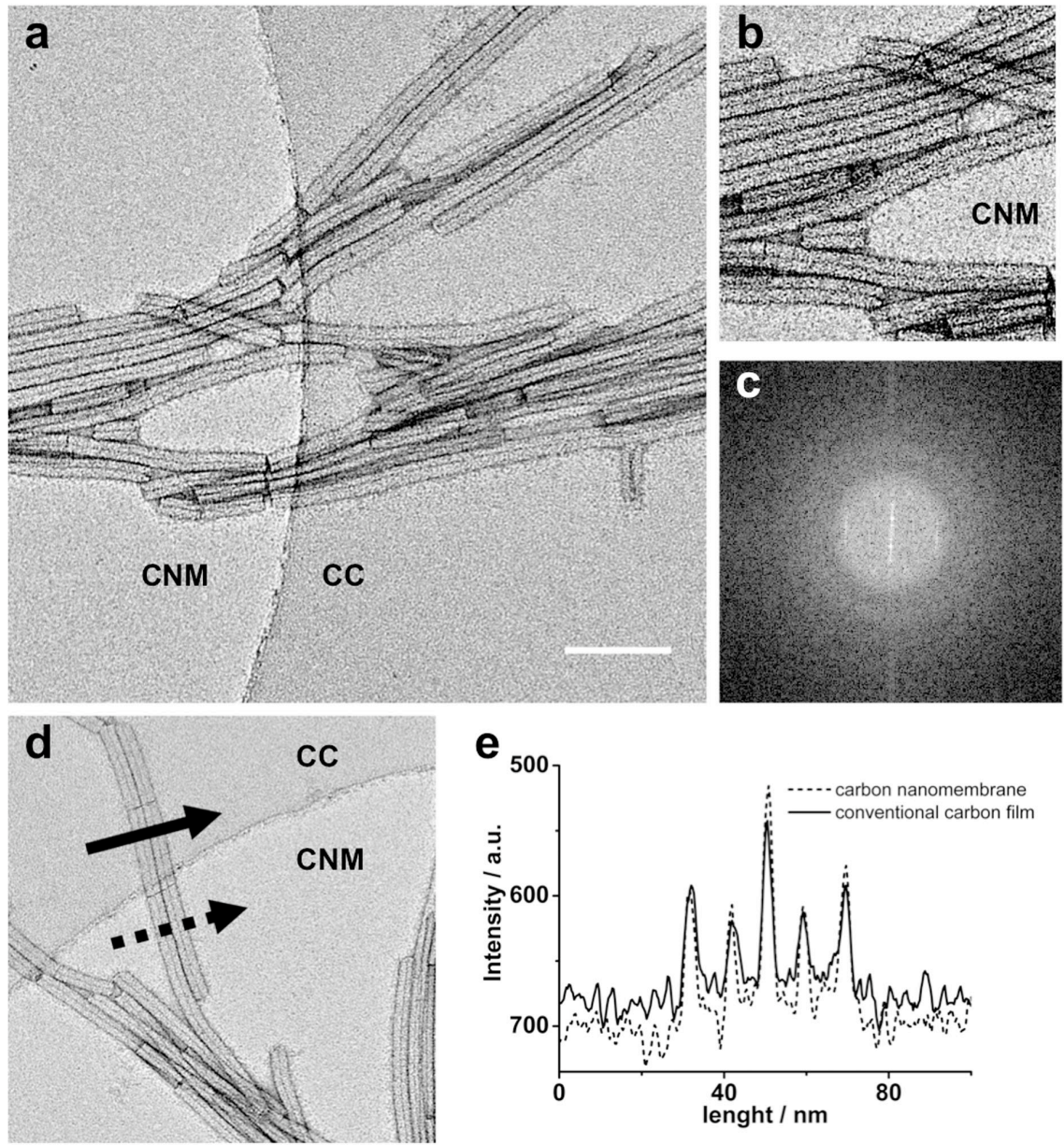

**Figure 1**

TEM images (non-filtered) of negatively stained TMV at room temperature on CNM support films. (a) The contrast of stained TMV on CNM is significantly higher compared to TMV on conventional carbon (CC). The scale bar is 100 nm. (b) Image and (c) corresponding Fourier transform of TMV on CNM. (d), (e) Image intensity measured along line sections perpendicular to TMV on CNM (dashed arrow) and CC (solid arrow) supports.

The Fourier transform (Fig. 1c), calculated from Fig. 1b, reveals the 3rd layer line of TMV. Line sections along both kinds of TMV (Fig. 1d) demonstrate that CNM considerably improve the contrast of TMV (Fig. 1e). We calculated a contrast gain of 28 % for negatively stained TMV on CNM compared to TMV on conventional carbon (CC). Fig. 2a shows zero-loss energy-filtered images of negatively stained TMV on CNM, imaged at cryogenic temperature. A uranium map of the same specimen area is shown in Fig. 2b. The corresponding uranium post-edge image (O edge, 96 eV) (Fig. 2c) demonstrates that the CNM support contributes virtually no background signal. Figure 2d shows a carbon map as obtained from the same area. The carbon signal of TMV interferes with the carbon signal of the CC support film. Within the same acquisition time the carbon signal of ~1 nm thin CNM is negligible compared to 18 nm thick TMV, thus enabling the acquisition of background-free carbon maps. We conclude that CNM are advantageous supports for EELS and ESI analysis of samples comprising heavy as well as light elements. The iron storage protein ferritin is abundant in eukaryotic and prokaryotic organisms and plays a major role in iron metabolism [26]. Ferritin comprises a protein shell and a ferrihydrite core, which has long been studied by electron microscopy and x-ray crystallography [27-30]. Beyond its biological function, ferritin has attracted attention as a building block for nanobiotechnology due to its ability of catalyzing the formation of metal nanoparticles within the protein shell [31, 32]. Using ferritin as a test specimen, we recorded EFTEM images of ferritin on CNM (Fig. 3a). The low granularity of CNM supports, compared to conventional amorphous carbon, facilitates high-resolution imaging of ferritin cores (Fig. 3b). CNM are more than six times thinner than the ferrihydrite core of ferritin (Fig. 3c), thereby allowing the acquisition of virtually background-free iron maps. Fig. 3e shows the iron post-edge image (M edge, 54 eV) of the same specimen area as in Fig. 3d. The conventional amorphous carbon produces a strong background signal whereas CNM produce no background signal, further demonstrating the usefulness of CNM support for ESI and EELS, in particular for biological samples.

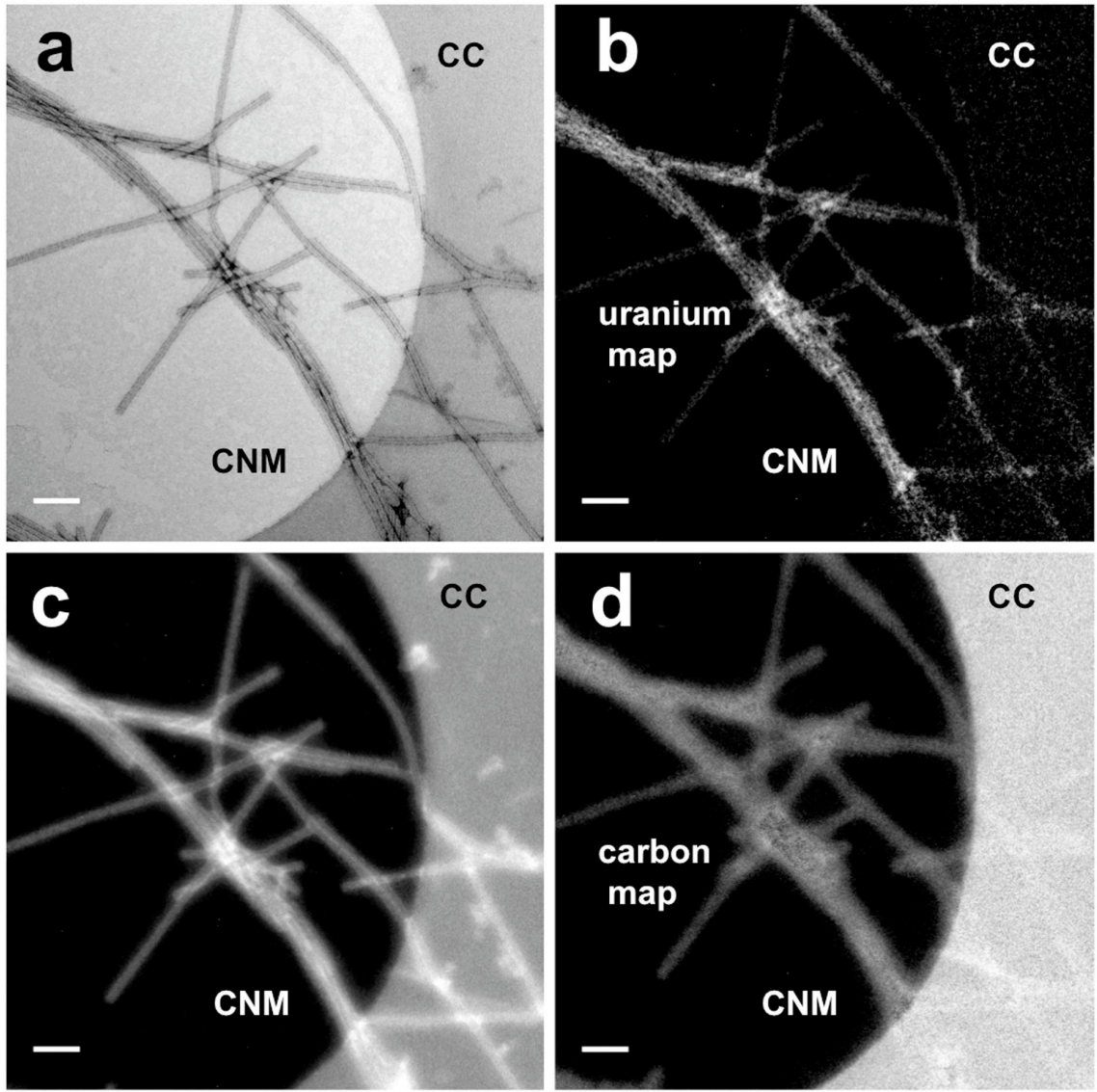

**Figure 2**

EFTEM of negatively stained TMV on CNM at liquid nitrogen temperature. (a) Zero-loss image of TMV on CNM. (b) Uranium map and (c) Uranium post-edge image (O edge, 96 eV) of the same specimen area. (d) Carbon map of the same specimen area. No background signal due to the CNM support is visible. Scale bars are 100 nm

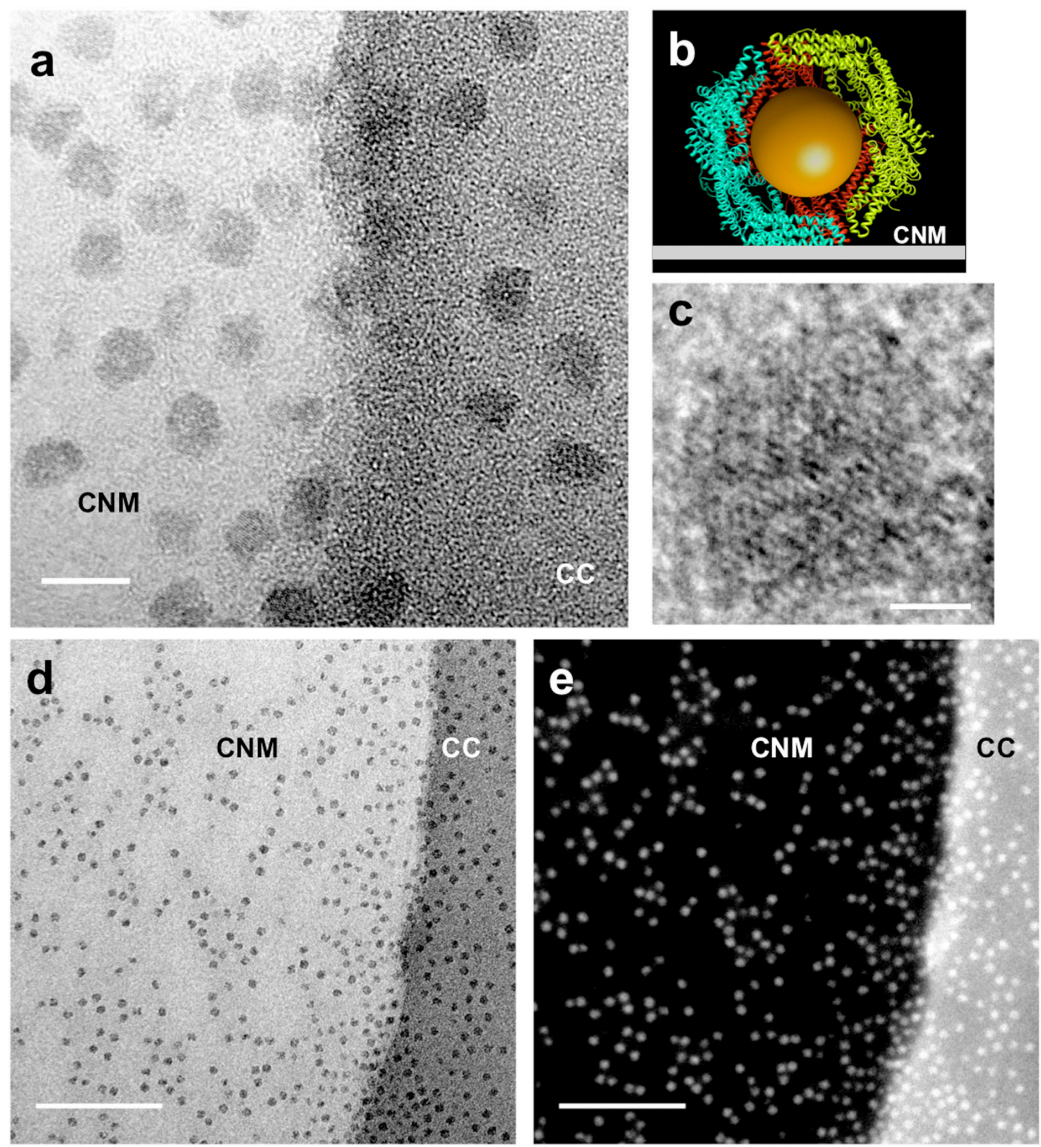

**Figure 3**

EFTEM images of ferritin on CNM recorded at liquid nitrogen temperature. (a) Zero-loss image of ferritin on freestanding CNM on holey carbon close to the edge of a hole. The scale bar is 10 nm. (b) Model of a single ferritin molecule on CNM, comprising a protein shell as well as a ferrihydrite core of ~ 6 nm diameter. Model created with Chimera [38]. Ferritin and CNM are drawn to scale. (c) High-resolution image of ferritin on CNM. The scale bar is 2 nm. (d) Zero-loss image of ferritin on CNM and (e) iron post-edge image (M edge, 54 eV) of the same area. In contrast to the CC support, no background signal from the CNM is detectable. Scale bars are 100 nm.

*3.2. EFTEM of ice-embedded specimens on conductive carbon nanomembranes (cCNM)*

Structure determination of biomolecules by single particle cryoEM is usually done with flash-frozen aqueous suspensions of macromolecular assemblies suspended over the holes of a holey carbon support film. However, it might be advantageous to include an additional thin support film for mechanical stability and electrical conductivity. With conventional holey carbon support films, particles very often accumulate at the edge of the hole, attach preferentially to the surrounding carbon film, or are removed from the aqueous film in the holes during washing steps. These problems can be overcome by inclusion of a conductive carbon nanomembrane (cCNM), obtained by thermal conversion of CNM to graphene [11], spanning the holes. The metallic nature of cCNM has been demonstrated in a previous work [11]. The sheet resistivity of cCNM reaches 100 kΩ/sq, which is just 15 times larger than the sheet resistivity of defect free graphene (6.5 kΩ/sq) [33]. A rough estimate for the bulk conductivity of cCNM is obtained by multiplying the sheet resistivity by the thickness of cCNM, which has been determined to 0.7 nm [11]. This gives a bulk resistivity of $7*10exp(-5)$ Ωm. For comparison, the bulk resistivity of metallic TiSi glasses has been determined to $1.46*10exp(-6)$ Ωm [6]. As vitrified water is an electrical insulator, an additional thin film of conducting material reduces beam-induced charging. Furthermore, cCNM add mechanical strength to the thin layer of vitreous buffer normally examined in cryoEM, which helps to reduce beam-induced specimen movements. We analyzed ice-embedded TMV on cCNM prepared using two different methods. In the first approach we vitrified TMV on top of a cCNM supported by lacey carbon (Fig. 4a). Figure 4b shows an overview image of vitrified TMV on cCNM. Within the specimen area shown the freestanding ~ 1 nm thin membrane spans a hole of ~ 6.8 μm diameter, thus demonstrating the high mechanical strength of cCNM. Fig. 5a shows a close-up view of ice-embedded TMV on cCNM. The Fourier transform (Fig. 5b) reveals the 3$^{rd}$ layer line of TMV indicative of 23 Å resolution. Proper hydration of the samples was verified by illuminating the same specimen area for a certain amount of time. After 20 s of illumination, corresponding to an electron dose of ~ 300 $e^-/Å^2$, characteristic bubbles due to water radiolysis [34] have formed along the virus particles, thus proving proper hydration (Fig. 5c). The Fourier transform (Fig. 5d) reveals stronger Thon rings, probably due to ice sublimation and sample decomposition. We analyzed the gain in contrast for ice-embedded TMV on ultrathin cCNM (Fig. 6a) compared to ice-embedded TMV on CC (Fig. 6c). The $3^{rd}$ layer line of TMV is visible in the Fourier transforms of both images (Fig. 6c,d). However, compared to the cCNM supports, stronger Thon rings due to the thick CC support film are visible in the case of TMV on CC (Fig. 6d). Fig. 6g shows line sections perpendicular to the axis of two selected virus particles imaged on cCNM (Fig. 6e) and CC (Fig. 6f), which demonstrate that the contrast of ice-embedded TMV is considerably higher on cCNM than on CC. We calculated a contrast gain of 35 % for vitrified TMV on cCNM compared to TMV on CC.

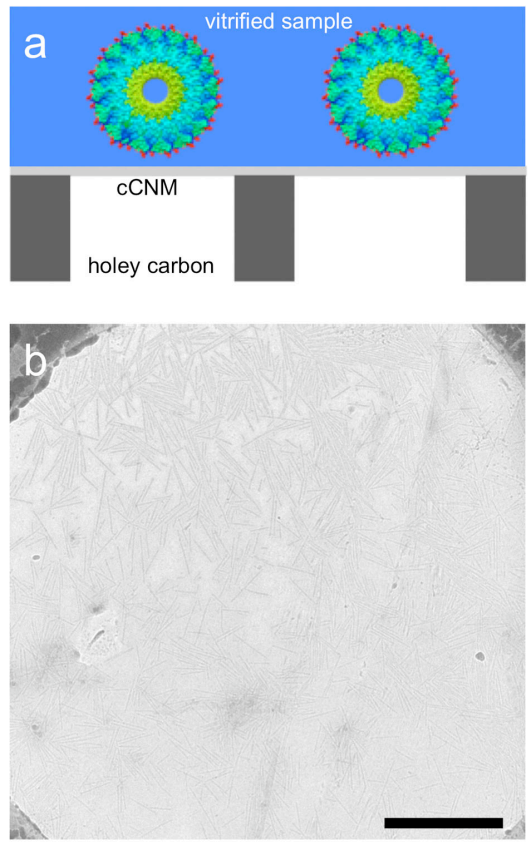

**Figure 4**

CryoEM of ice-embedded TMV on cCNM. (a) Specimen preparation method. TMV is vitrified on top of a continuous cCNM supported by conventional lacey carbon. (b) Overview image (non-filtered). Shown is vitrified TMV on a freestanding cCNM spanning a hole of ~ 6.8 μm diameter. The scale bar is 1 μm.

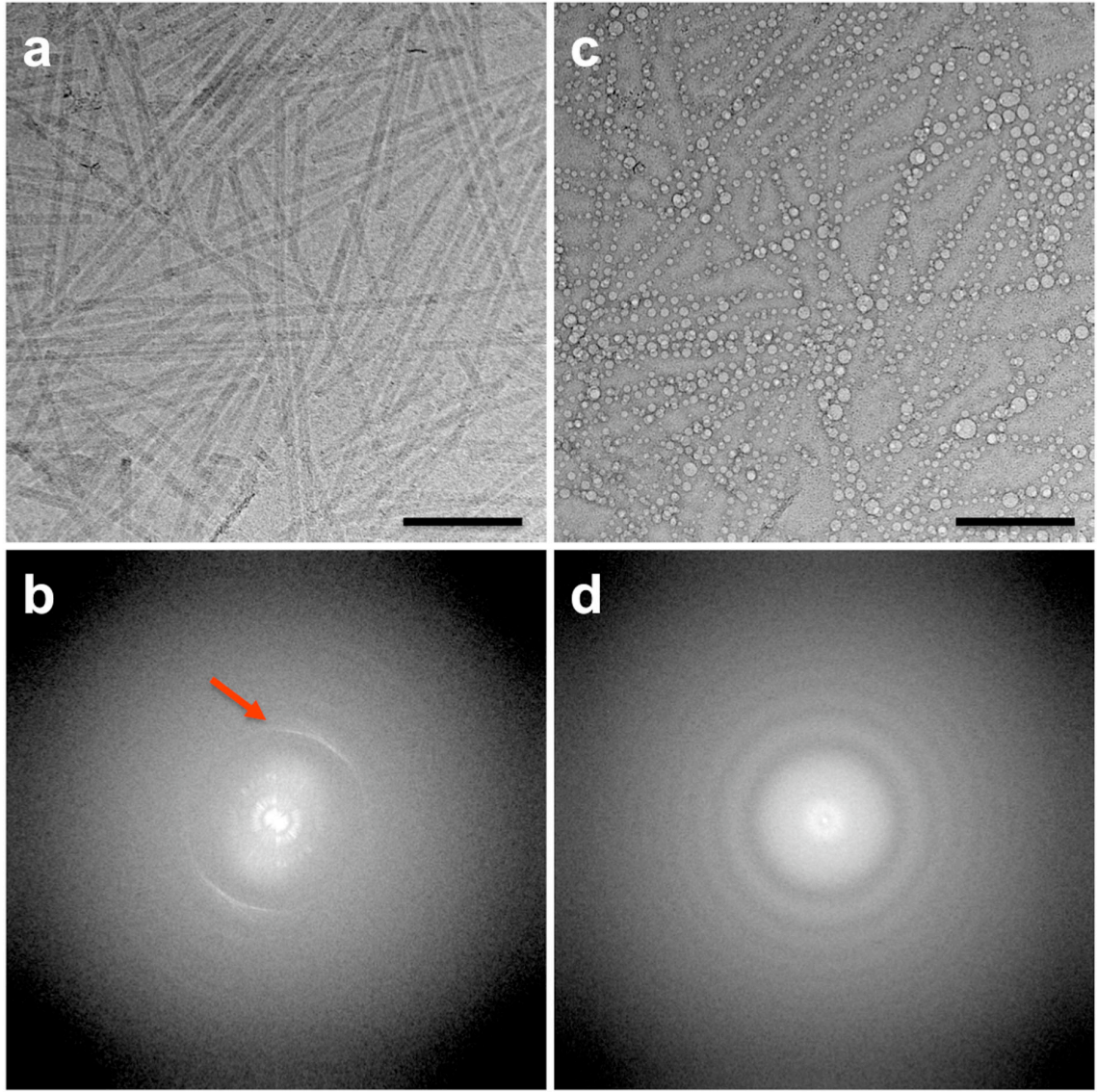

**Figure 5**

Hydration of vitrified TMV on cCNM. (a) Image (non-filtered) of ice-embedded TMV on cCNM. Scale bar is 100 nm (b) FFT of (a). The 3$^{rd}$ layer line of TMV is visible (red arrow). (c) The same specimen area after 20s illumination. Characteristic bubbles along the virus particles are visible indicative of a hydrated sample. Scale bar is 100 nm (d) FFT of (c).

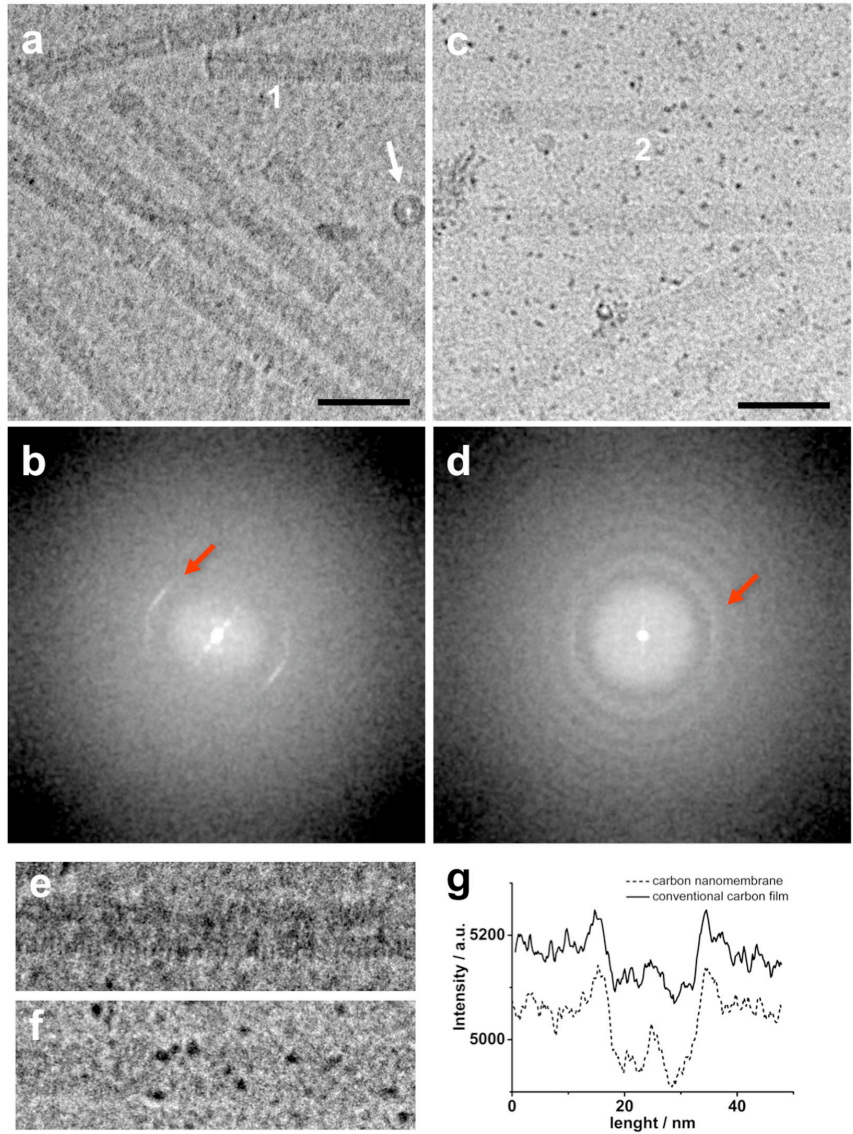

**Figure 6**

Contrast quantification (a) Ice-embedded TMV on cCNM with end-on view of a virus fragment (white arrow) (b) FFT of (a). The 3$^{rd}$ layer line of TMV is indicated (red arrow). (c) Ice-embedded TMV on CC. (d) FFT of (c). The 3$^{rd}$ layer line of TMV is marked by red arrow. (e) Ice-embedded virus particle on cCNM used for contrast quantification (labeled 1 in (a)). (f) Ice-embedded virus particle on CC used for contrast quantification (labeled 2 in (c)). (g) Line sections of (e) and (f) perpendicular to the virus axis. Accounting for the respective background intensities, the image contrast of TMV on cCNM is higher than the contrast of TMV on CC.

The second method we have developed for the preparation of ice-embedded specimens on cCNM is sketched in fig. 7a. A grid comprising cCNM on holey carbon is loaded with the aqueous sample from the carbon facing side of the cCNM. Afterwards the grid is blotted and plunged into liquid ethane, producing ice-embedded specimens supported by cCNM. Figure 7b shows an EFTEM image of ice-embedded TMV samples prepared by this method. The Fourier transform (Fig. 7c) reveals the $3^{rd}$ and $6^{th}$ layer lines of TMV, indicative of 11.5 Å resolution. Apparently a thin layer of water remains strongly bound in the cavities formed by holes and cCNM due to capillary forces, so that the specimen is fully embedded in a thin layer of vitreous ice. Indeed, we observed that even samples blotted extensively for more than 5 s were still fully hydrated. Compared to preparation of ice-embedded specimens on continuous support films, this preparation method avoids artifacts due to undesired strong sample-substrate interactions or compression of the sample at the water-air interface. Avoiding both sources of artifacts is a prerequisite for single particle cryoEM at near-atomic resolution [35]. Combining standard holey carbon and ultrathin conductive supports (cCNM, graphene) might become a generally useful technique to improve specimen preparation for single particle cryoEM.

**Conclusions**

In conclusion, we have demonstrated that energy-filtered TEM of biological samples benefits from using ~1 nm thin CNM support films. Due to their exceptional thinness CNM considerably improve the signal-to-noise ratio of energy-filtered TEM images, thus advantageous for ESI and EELS of biological samples, which comprise mainly light elements. Beyond their application in biological EFTEM ultrathin CNM probably will be useful as support films for energy-filtered TEM and STEM of inorganic materials as well, i. e. for high-resolution plasmon EELS of metallic nanoparticles [36, 37].
Furthermore, we have shown that cCNM are suitable supports for cryoEM of frozen-hydrated samples. The preparation methods presented lead to frozen-hydrated specimens suspended over holes, supported by an ultrathin conductive film. While resembling graphene with respect to conductivity and mechanical strength, continuous cCNM can be produced at any size as desired, i. e. with dimensions matching the size of an EM grid.

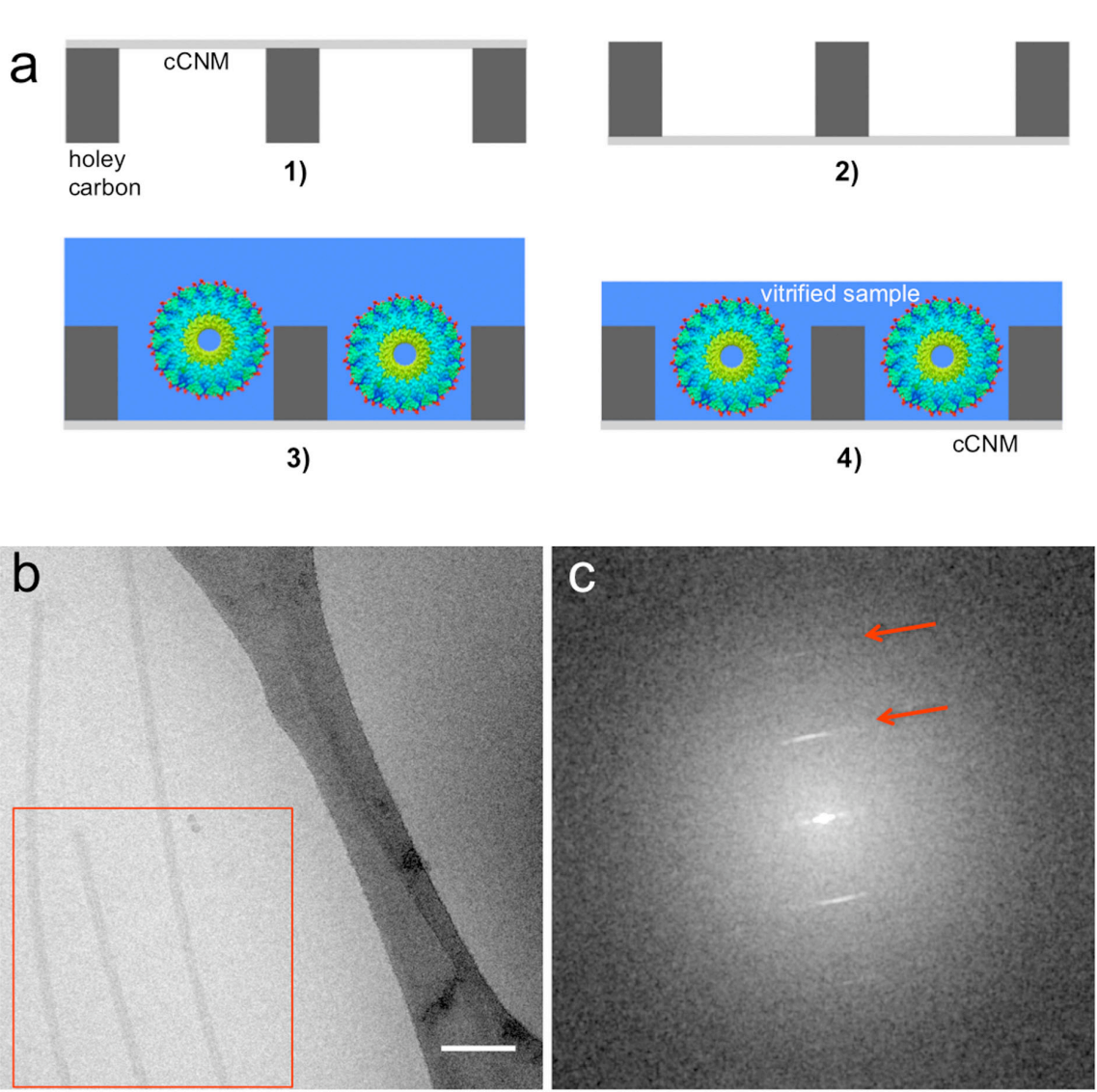

**Figure 7**

EFTEM imaging of ice-embedded TMV supported on ~1 nm thin cCNM. (a) Specimen preparation method. A grid coated with cCNM on holey carbon (1) is loaded from the carbon facing side of the cCNM (2, 3). The grid is blotted and plunged into liquid ethane (4). (b) EFTEM CCD image of ice-embedded TMV on cCNM prepared according to (a). (c) Corresponding Fourier transform (area indicated in (b)). The 3$^{rd}$ and 6$^{th}$ layer lines of TMV are visible (arrows).